\documentclass[%
 reprint,
 amsmath,amssymb,
 aps,
]{revtex4-2}

\usepackage{graphicx}
\usepackage{dcolumn}
\usepackage{bm}
\usepackage{subfigure,amsfonts}

\begin{document}

\preprint{APS/123-QED}

\title{Ultrafast non-equilibrium magnon generation and \\ collapse of spin-orbit hybridization gaps in iron}

\author{Xinwei Zheng}%
\affiliation{Fachbereich Physik, Freie Universität Berlin, Arnimallee 14, 14195 Berlin}%

\author{Shabnam Haque}%
\affiliation{Fachbereich Physik, Freie Universität Berlin, Arnimallee 14, 14195 Berlin}%

\author{Christian Strüber}%
\affiliation{Fachbereich Physik, Freie Universität Berlin, Arnimallee 14, 14195 Berlin}

\author{Martin Weinelt}%
\email{martin.weinelt@fu-berlin.de}
\affiliation{Fachbereich Physik, Freie Universität Berlin, Arnimallee 14, 14195 Berlin}%
\homepage{http://www.physik.fu-berlin.de/en/einrichtungen/ag/ag-weinelt/}

\date{\today}

\begin{abstract}
We distinguish between longitudinal and transverse spin excitations in the ultrafast response of iron by probing exchange splitting $\Delta E_{\rm{ex}}$ and magnetic linear dichroism (MLD) in time- and angle-resolved photoemission. Comparing spin-split partner bands at the Fermi level shows that $\Delta E_{\rm{ex}}$ remains constant upon optical excitation. In contrast, the different MLD response of spin-orbit-split valence bands reveals non-equilibrium, transverse spin dynamics. Magnon generation in Fe is ultrafast, electronic band specific, and drives the collapse of spin-orbit hybridization gaps.
\end{abstract}


\maketitle

Excitation with optical femtosecond (fs) pulses brings magnetically ordered systems far out of thermal equilibrium and opens up alternative paths of controlling spin dynamics. In metallic ferromagnets, the electronic system is heated to few thousand Kelvin and thus transfers energy to spin and phonon degrees of freedom. The question of how electrons couple to the spin system is extremely important, but our microscopic understanding is far from complete. Is the magnetic moment per atom primarily reduced or remain the magnetic moments constant but become mutually tilted? While the former longitudinal spin excitation manifests in a decrease of the exchange splitting, the latter transverse spin excitation is accompanied by a drop in spin polarization \cite{Weissenhofer2025}. There is evidence for longitudinal \cite{Rhie2003,Koopmans2010,Mueller2013,Griepe2023} and transverse \cite{Carpene2008,Schmidt2010,Carpene2015,Eich2017,Gort2018} spin excitations or both \cite{Tengdin2018}, reflecting the controversial findings. 
To this end the $3d$ metals are regarded as prototype systems for the interplay of band magnetism and correlations \cite{Sanchez-Barriga2012,Tusche2018}. 
With a high density of states at the Fermi level $E_{\rm{F}}$, the $3d$ itinerant ferromagnets fulfill the Stoner criterion and show in thermal equilibrium predominantly a reduction of the magnetic moment due to a decrease of the exchange splitting (Stoner behavior) \cite{Greber1997}. However, this can be significantly different out of thermal equilibrium. Higher order terms in electron-electron interaction and non-local correlations open the path to ultrafast magnon generation by inelastic scattering of optically excited electrons, \textit{i.e.} already in the first 200\,fs during internal thermalization of the electron subsystem \cite{Bovensiepen2007,Weinelt2025}. 
Ab-initio calculations showed a strong influence of magnon emission on the lifetime of excited minority spin carriers in Fe, while this effect was predicted to be negligible in Ni \cite{Zhukov2004}. The former was corroborated by studying the spin-dependent lifetime of image-potential states on Fe(001) where magnon emission doubles the phase space for inelastic scattering of minority electrons \cite{Schmidt2010}. The effect is less pronounced for Co(001) \cite{Weinelt2025}, which indicates an increasing influence of magnon generation on ultrafast spin dynamics of Ni to Co and Fe. 
In line, initial time- and angle-resolved photoelectron spectroscopy (tr-ARPES) studies on Ni report on a collapse of the exchange splitting within 300\,fs attributed to spin flip scattering, i.e. longitudinal spin excitations \cite{Rhie2003}. Later, the simultaneous heating of the spin system was attributed to transverse spin excitations \cite{Tengdin2018}. In contrast for Co and Fe, spin-resolved ARPES data could be well reproduced assuming a rigid but spin-mixed band structure \cite{Eich2017,Gort2018}. Measuring the spin polarization is very demanding. Therefore the latter experiments were performed in normal emission at fixed photon energy and probe a small, rather unspecific section of the bulk band-structure.   

\begin{figure}[b]
    \centering
    \includegraphics[width=1\linewidth]{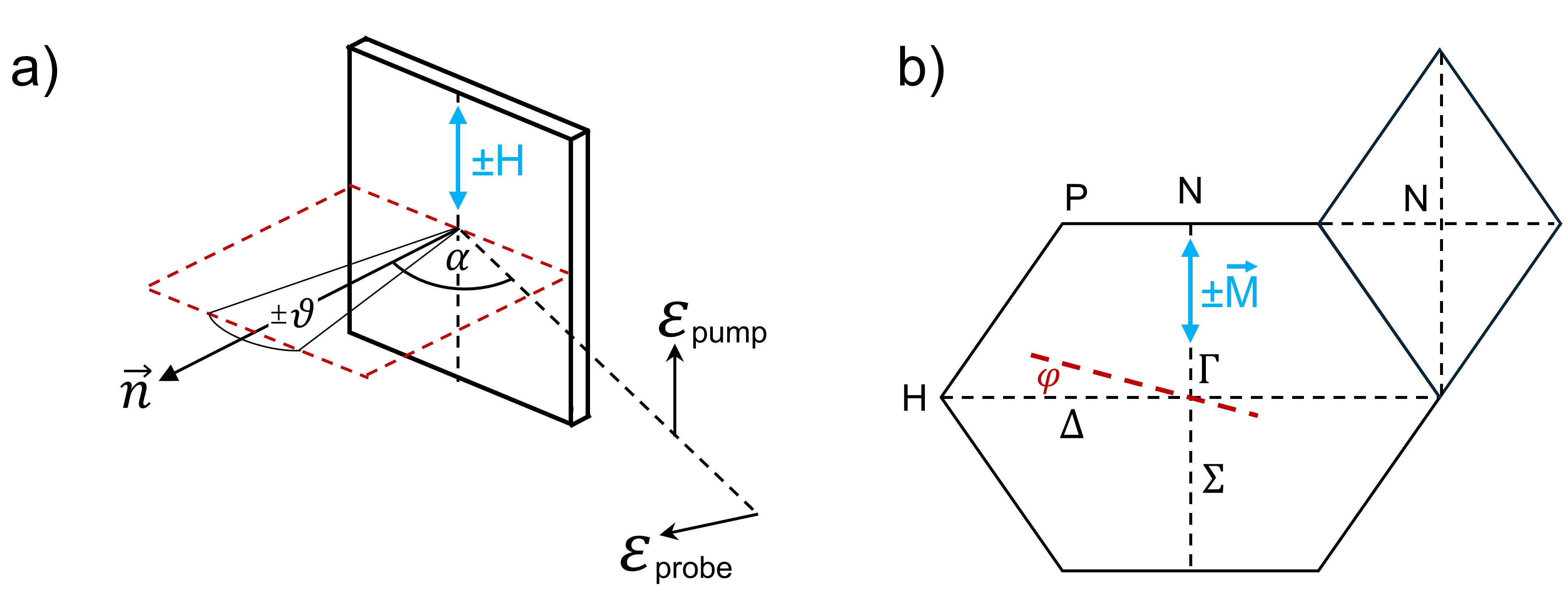}
    \caption{a) Measurement geometry for tr-ARPES. b) Section of the $\Gamma$HN-plane of the bulk Brillouin-zone (BZ). The easy axis of magnetization $\pm \vec{M}$ lies in-plane parallel to the $[1\overline10]$ direction ($\Gamma$-N). The band dispersion $E(k_\|)$ is measured for parallel momenta $k_\|$ indicated by the dashed red line.} 
    \label{fig:geometry}
\end{figure}

In the present work, we demonstrate an alternative approach to distinguish between longitudinal and transverse spin excitations following the exchange splitting $\Delta E_{\rm{ex}}$ and the transient magnetic linear dichroism (MLD) of the valence bands in tr-ARPES, respectively. We probe a pair of exchange-split $d$-bands that intersect the Fermi level at parallel momenta $k_\parallel = 0.3$ and 1.4\,{\AA}$^{-1}$ along the \mbox{$\Gamma$-H} direction in the 2nd Brillouin zone (BZ). $\Delta E_{\rm{ex}}$  remains unchanged upon optical excitation. In addition, we follow the MLD contrast of spin-orbit-split 
minority spin $d$-bands around $\Gamma$. The decay of the MLD contrast reveals non-equilibrium transverse spin dynamics causing ultrafast spin mixing within 150 fs. Spin mixing leads to a collapse of the spin-orbit hybridization gap and a concomitant $\sim 40$\,meV change in band positions.  

\begin{figure*}[t]
    \centering
\includegraphics[width=0.8\textwidth]{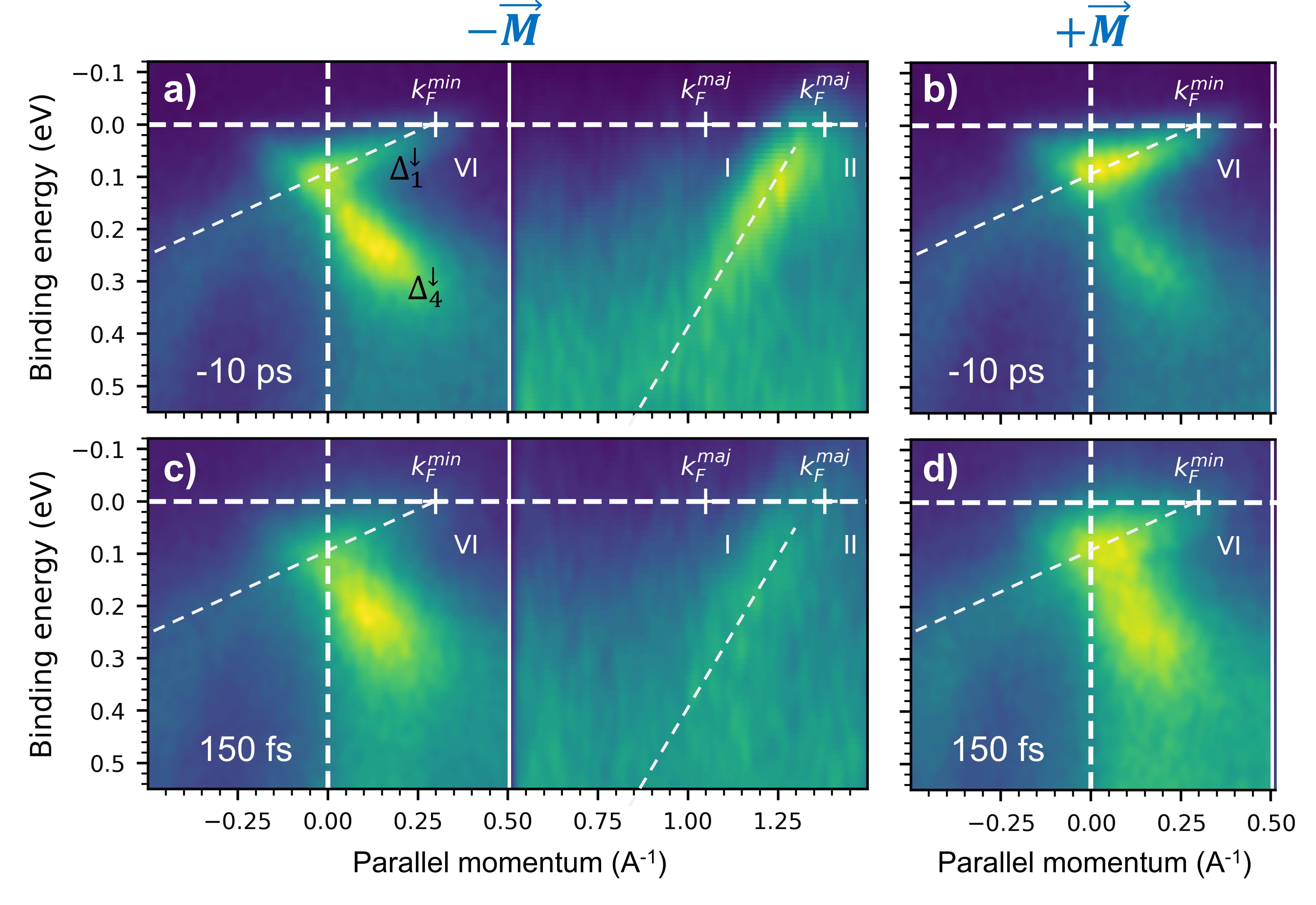}
\caption{Band dispersion $E(k_\|)$ recorded for a) -10\,ps delay, $-\bf{M}$, b) -10\,ps delay $+\bf{M}$ c) 150\,fs delay, $-\bf{M}$ and d) 150\,fs delay, $+\bf{M}$. For negative pump-probe delay the pump pulse arrives after the probe pulse, i.e. comparable to the non-pumped, static case (\textit{cf} SM, Figs.~S2). The measurement directions $k_\|$ is indicated by the red dashed line in d). The photon energy of IR pump and VUV probe pulses was 1.58 and 26.9\,eV. The absorbed pump fluence was $1.9\,\rm{mJ/cm}^2$. The exchange-split bands of minority spin electrons (VI) and of the corresponding majority spin electrons (II) intersect $E_{\rm{F}}$ at about $k_\parallel = 0.3$ and 1.4\,{\AA}$^{-1}$. Since we do not observe MLD for $k_{\|} \geq 0.5$\,\AA$^{-1}$ (see SM, Fig.~S2), we did not record spectra for $+\bf{M}$.} 
\label{fig:band_structure}
\end{figure*}
 
The experimental setup for tr-ARPES and higher harmonic generation (HHG) is described in Ref.\,\onlinecite{Frietsch2013}. The 15 monolayer (ML) single-crystalline Fe(110) film was prepared via molecular beam epitaxy on a W(110) substrate at 300\,K and a pressure of $3 \cdot 10^{-10}$\,mbar during evaporation. After annealing to 530\,K the Fe film showed a sharp low-energy electron diffraction pattern typical for the body-centered cubic (bcc) (110)-surface \cite{Gradmann1982,Andres2015}.\\ 
tr-ARPES measurements were performed at 100\,K sample temperature and a base pressure of $5 \cdot 10^{-11}$\,mbar. The geometry of the experiment is sketched in Fig.~\ref{fig:geometry}a. The vacuum-ultraviolet (VUV) probe pulses with a photon energy of 26.9\,eV (17th harmonics of the 1.58\,eV laser fundamental) were $p$-polarized and impinged at an angle of $\alpha = 60^{\circ}$ with respect to the surface normal $\vec{n}$. Photoelectrons were detected after a hemispherical photoelectron analyzer with acceptance angle of $\pm 15^{\circ}$ in an energy range of 2.5\,eV below the Fermi level $E_{\rm{F}}$ in the plane spanned by $\vec{n}$ and the electric field vector $\vec{\cal{E}}_{\rm{probe}}$ of the probe pulse. The Fe film was magnetized in-plane ($\pm \bf{M}$ parallel to the $[1\overline{1}0]$-direction, Fig.~\ref{fig:geometry}b) by a magnetic field pulse of 100\,mT using a free-standing coil.\\ 
The infrared (IR) 1.58-eV pump pulse impinged nearly collinear to the probe pulse. The pulse duration and absorbed fluence (in Fe and W) were 50\,fs and 1.9\,mJ/cm$^2$, respectively. To reduce space charge by multi-photon photoemission the pump pulses was $s$-polarized. The 10\,kHz repetition rate of the laser amplifier allowed for data acquisition in counting mode ($\sim 0.5$ counts per laser pulse). The overall time and energy resolution of the tr-ARPES experiment were 160\,fs and 80\,meV.

As reference for our study, we used the band-structure and Fermi-surface measurement of Fe(110) by Sch{\"a}fer and coworkers \cite{Schaefer2005}. They tuned the photon energy and observed the $\Gamma$ point of the 3rd BZ at $h\nu = 136.3$\,eV in normal emission along $\Gamma$-N. The perpendicular moment amounts to $k_\bot = 0.511 \cdot \sqrt{h\nu + V_0}$\,\AA$^{-1}$, where $V_0$\,(in eV) is the muffin-tin zero referenced to $E_{\rm{F}}$. With an extent of the BZ of $2\cdot\overline{\Gamma{\rm{N}}} = 3.096$\,{\AA}$^{-1}$ this results in an inner potential of $V_0 = 9.8$\,eV. Assuming $V_0$ to be independent of photon energy, $h\nu = 26.9$\,eV corresponds to $k_\bot =  3.095$\,\AA$^{-1}$. Thus by selecting the 17th harmonics of the HHG source we probe the band structure close to the $\Gamma$ point of the 2nd BZ for $k_\|$ in the $\Gamma\rm{HN}$ - plane (see Fig.~\ref{fig:geometry}b and Fig.~S1 in the Supplemental Material, SM).

Figure\,\ref{fig:band_structure} shows the corresponding $E(k_{\parallel})$ dispersion for magnetization $-\bf{M}$ at a) -10\,ps  and c) 150\,fs pump-probe delay and for $+\bf{M}$ at b) -10 ps and d) 150\,fs delay. For negative delay the IR pump pulse arrives after photoemission by the VUV probe pulse. Corresponding static measurements without pump pulse are shown in SM, Fig.~S2. We measured a path $15^{\circ}$ rotated to the $\Gamma$-H direction (dashed red line in Fig.~\ref{fig:geometry}b) and (with the exception of Fig.~\ref{fig:band_structure}d) merged two data sets, which were recorded $\pm 15^{\circ}$ around polar emission angles of $\vartheta = 0^{\circ}$ and $25^{\circ}$. The data compare well with those recorded in the 3rd BZ along $\Gamma$-H (\textit{cf}~Fig.\,12 in Ref.\,\onlinecite{Schaefer2005}). The Roman numbers denote the Fermi sheets in the (110) plane following Sch{\"a}fer \textit{et al.} There is a static shift of all bands to lower binding energy $E_{\rm{B}}$ by about 100\,meV (see also SM, Fig.~S2) which we attribute to a confinement effect in the 15\,ML film \cite{Andres2022} as compared to the thicker 100\,ML film studied in Ref.\,\onlinecite{Schaefer2005}. Strong correlations lead to electron lifetimes of $\leq 15$\,fs (200\,meV) at $E - E_{\rm{F}} \geq 300$\,meV \cite{Weinelt2025}, which likewise explains extreme blurring of the bands at $E_{\rm{B}} \geq 300$\,meV \cite{Sanchez-Barriga2012}.\\ 

The bands VI and II correspond to exchange-split minority and majority spin partner bands (\textit{cf}  band-structure calculation in Fig.\,7 of Ref.\,\onlinecite{Schaefer2005}), which intersect the Fermi level at about $k_{\rm{F}}^{\rm{min}} = 0.3$\,{\AA}$^{-1}$ and 
$k_{\rm{F}}^{\rm{maj}} = 1.4$\,{\AA}$^{-1}$. Obviously, it is essential to probe these spin-partner bands close to $E_{\rm{F}}$ to determine the transient exchange splitting. Upon optical excitation, one would expect an upwards and downwards shift of the minority and majority spin bands, respectively and a corresponding reduction in their $k_{\|}$ distance. The dashed white lines in Fig.~\ref{fig:band_structure} are linear approximations to the band dispersion close to $E_{\rm{F}}$. Comparing the $E(k_\|)$ maps in Figs.~\ref{fig:band_structure} we see significant intensity changes but no obvious shift of the bands upon IR excitation. For a comparison of energy distribution curves of the bands close to the Fermi level at $k_F^{\rm{min}}$ and $k_F^{\rm{max}}$ we refer to SM, Fig.~S3. We conclude that the maximum reduction of the exchange splitting is below $20$\,meV, which corresponds to about 1\% of the Fe exchange splitting. Cum grano salis, the band structure remains rigid upon IR excitation and the number of majority and minority spin electrons constant. We conclude that longitudinal spin excitations are negligible (for an $E(k_\|)$ map at 1 ps delay see SM, Fig.~S4).

Pump-probe experiments recording the magneto-optical Kerr effect (MOKE) showed a clear decrease of the MOKE contrast for Fe. This was assigned to the generation of magnons \cite{Carpene2008,Carpene2015}. Since the origin of MOKE is spin-orbit coupling (SOC), we expect a similar dynamical response when studying MLD in the angular distribution of photoelectrons. For a detailed experimental and theoretical study of the MLD on Fe(110) we refer to Ref.~\onlinecite{Rampe1998}. Rampe \textit{et al.} recorded and simulated photo\-emission spectra in normal emission ($k_{\|} = 0$) as a function of photon energy probing MLD along $k_\bot$, i.e. $\Gamma$-N. 

\begin{figure}[t]
    \centering
    \includegraphics[width=1.0\linewidth]{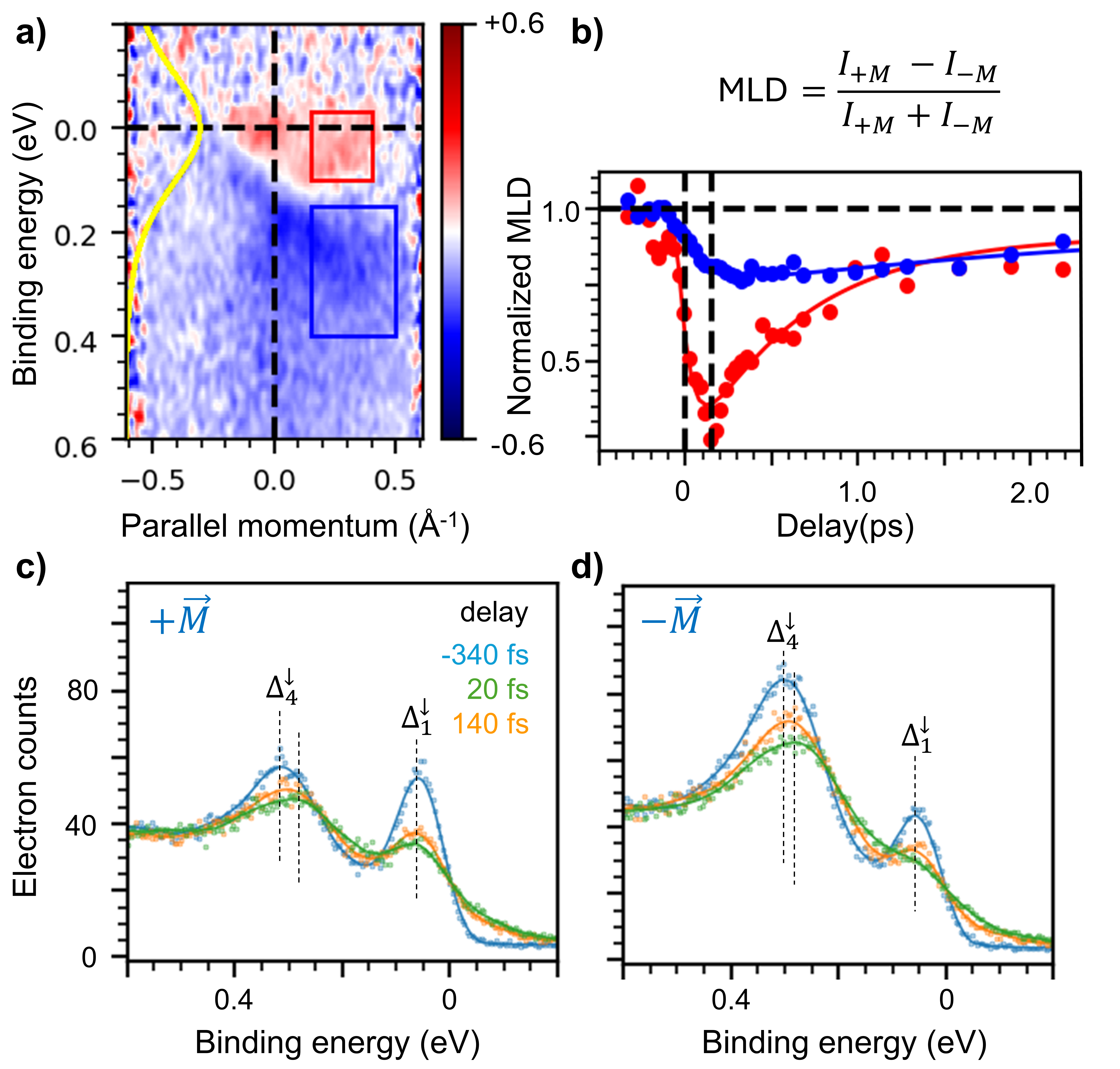}
    \caption{a) Static MLD as a function of binding energy and parallel momentum. The phase space for electron scattering around $E_{\rm{F}}$ is indicated by the left, yellow solid line. b) MLD as a function of pump-probe delay. Data have been averaged over the red and blue areas in Fig.~\ref{fig:mld}a and are normalized to the static MLD signals. Error bars are within symbol size. Solid lines are guides to the eye from bi-exponential fits.
    Energy distribution curves for opposite magnetization, C) $+\bf{M}$ and d) $-\bf{M}$ showing the $\Delta_1^\downarrow$\,- and $\Delta_4^\downarrow$\,-like minority spin valence bands extracted in a momentum range of $k_{\|} = 0.1 - 0.2$\,\AA$^{-1}$ for pump-probe delays of -340, 20, and 140\,fs.}
    \label{fig:mld}
\end{figure}

Figure~\ref{fig:mld}a shows the MLD map $(I_{+{\bf{M}}} - I_{-{\bf{M}}})/(I_{+{\bf{M}}} + I_{-{\bf{M}}})$ recorded for polar angles  $\vartheta = 0 \pm 15^{\circ}$ and $h\nu = 26.9$\,eV. $I_{\pm {\bf{M}}}$ corresponds to the photoemission intensity maps measured for opposite magnetization directions $\pm\bf{M}$ (see Figs.~\ref{fig:band_structure}a, b and S2a, b in SM). We observe a dichroic contrast of about 15\% in a parallel momentum range of $-0.1\,\rm{\AA}^{-1} \leq k_\| \leq 0.5\,\rm{\AA}^{-1}$. As seen from Figs.~\ref{fig:band_structure}a and b the MLD signal is related to the upward and downward dispersing $\Delta_1^{\downarrow}$\,- and $\Delta_4^{\downarrow}$\,-like spin minority bands merging at $k_{\|} = 0$ \cite{Note1}. The dichroic contrast shows opposite sign when comparing both bands. We use this contrast to follow the magnetization dynamics. Figure~\ref{fig:mld}b depicts the MLD signal for the $\Delta_1^{\downarrow}$\,- and $\Delta_4^{\downarrow}$\,-like bands (red and blue circles) as a function of pump-probe delay. Data have been averaged over the squares in Fig.~\ref{fig:mld}b and are normalized to the corresponding values of the static (non-pumped) spectrum. We observe an ultrafast drop of the MLD signal within our experimental time resolution of 150\,fs and its recovery on a slower picosecond timescale. Both signals merge at around 1\,ps but show significantly different decay amplitudes at early times. At a delay of 1\,ps, we expect that equilibrium between electron, phonon and spin subsystems is reached. Consequently the MLD signals match and the sample recovers via cooling. To understand the non-equilibrium spin dynamics at early times we first need to understand the origin of MLD. 

Generally, in valence band photoemission dichroic signals are based on an interference effect between different initial states  \cite{Kuch2001}. The magnetization $\pm\bf{M}$ along $\Gamma$-N in the $\Gamma\rm{HN}$-plane (see Fig.~\ref{fig:geometry}b) lifts the mirror symmetry with respect to this plane, since $\bf{M}$ is an axial vector. Consequently, the bands distinguished by this mirror symmetry for $\bf{M} = 0$ become degenerate in the ferromagnetic state ($\bf{M} \neq 0$). For bcc iron this is, \textit{e.g.}, true for the $\Sigma_1^{\downarrow}$ and $\Sigma_3^{\downarrow}$ bands probed along the $\Gamma$-N-direction ($\Sigma$) \cite{Rampe1998,Schaefer2005} and in our study for the $\Delta_1^{\downarrow}$ and $\Delta_4^{\downarrow}$ bands probed along the $\Gamma$-H-direction ($\Delta$). In the paramagnetic state these bands merge at the $\Gamma$ point \cite{Schaefer2005}. In the ferromagnetic state the symmetry is lowered, the bands hybridize and a band gap opens. The driving mechanism of this avoided crossing is spin-orbit coupling, obviously absent for ${\bf{M}} = 0$. SOC leads in general to hybridization of both majority and minority spin bands, but for our case the contribution of majority spin bands is minor, since the Fe bands show a large exchange splitting of $\Delta_{ex} \sim 2$\,eV. The SOC-induced gap close to the $\Gamma$ point along $\Sigma$ was calculated to 70\,meV \cite{Rampe1998}.       
Since the photoemission matrix elements ${\cal{M}}_{if}$ contain a linear combination of initial states $i$, the photoemission intensity $\propto |{\cal{M}}_{if}|^2$ is affected by the corresponding interference terms. When reverting the direction of magnetization the contribution of initial states to the upper and lower band edge changes leading to different photoemission intensities $I_{\pm \bf{M}}$ and resulting MLD contrast $\propto I_{+\bf{M}} - I_{-\bf{M}}$. Considering just simplified linear combinations like $\alpha^\pm\cdot\Delta_1 \pm \beta^\pm\cdot\Delta_4$ can explain the sign change of the dichroic contrast between upper and lower band in Fig.~\ref{fig:mld}a. Obviously the hybridization of the initial states is not only strong at $\Gamma$ but there is still finite hybridization along the dispersing bands in the $\Gamma\rm{NH}$-plane up to $k_{\|} \leq 0.3$\,{\AA}$^{-1}$ \cite{Rampe1998}. Note that the width of the dichroic regions in Fig.~\ref{fig:mld}a is also caused by the large lifetime broadening of the valence band photoholes. 

The different dynamics of the MLD signals on the sub-ps timescale in Fig.~\ref{fig:mld}b can now be explained by an individual response of the two bands that hybridize. In other words, the delay-dependent spin mixing must be different for the two bands in the sub-picosecond regime. The band closer to the Fermi level shows a three times larger response. This indicates that the phase space available for inelastic electron scattering is relevant. Even at early times the hot non-equilibrium electron distribution can be roughly approximated by a Fermi distribution $f(E,T_{\rm{e}})$ at elevated electron temperature $T_{\rm{e}}$ \cite{Bovensiepen2007,Tengdin2018}. The electronic phase space for electron-magnon scattering is proportional to $f\cdot(1-f)$. From at fit of a Fermi distribution to the photoemission data we obtain a maximal electron temperature $T_{\rm{e}}$ of about 900\,K at 100\,fs delay. The corresponding phase space is indicated in Fig.~\ref{fig:mld}a by the yellow solid line. Clearly at early delays electron-magnon scattering will dominate electron dynamics at binding energies of $\pm 150$\,meV around $E_{\rm{F}}$. We conclude that this drives band-specific spin mixing.

What is more, ultrafast demagnetization should lead to an ultrafast drop in SOC. Figures~\ref{fig:mld}c and d show electron distribution curves (EDCs) displaying the $\Delta_1^\downarrow$\,- and $\Delta_4^\downarrow$\,-like bands extracted in a momentum range of $k_{\|} = 0.1 - 0.2$\,\AA$^{-1}$ for pump-probe delays of -340, 20, and 140\,fs. Comparing the EDCs for $+\bf{M}$ and $-\bf{M}$ we clearly recognize the MLD and its sign change between the two bands. While we hardly see a peak shift of the $\Delta_1^\downarrow$ band, the $\Delta_4^\downarrow$ band shifts by $40 \pm 10$\,\,meV to lower binding energy.  However, the minority spin bands should shift to a higher $E_{\rm{B}}$ when the exchange splitting $\Delta_{\rm{ex}}$ decreases. The observed shift to lower $E_{\rm{B}}$ is only compatible with a decrease of spin-orbit coupling and the concomitant collapse of the hybridization gap. We note that the hybridization gap opens close to the $\Gamma$ point and was calculated to 70\,meV \cite{Rampe1998}. We can not resolve this gap but still observe a 40\,meV shift at $k_{\|} = 0.15$\,\AA$^{-1}$. Moreover, the fact that the $\Delta_1^\downarrow$ band barely shifts to a lower binding energy again indicates that the two bands hybridize differently and that the contribution of a reduction in exchange splitting to the overall band shift is negligible. As discussed above, the bands are rigid with respect to exchange and only the ultrafast closing of the SOC-induced gap leads to the expected, albeit small change in binding energy.

In summary, we probe the band structure of Fe(110) for parallel momenta close to the $\Delta$ direction of the 2nd BZ with tr-ARPES. Upon optical excitation with a 50\,fs IR pulse we observe an ultrafast decay of the MLD signal within 150\,fs. In contrast the exchange splitting remains constant. Thus, Fe demagnetizes via ultrafast excitation of transverse spin excitations generated by inelastic electron-magnon scattering of the optically excited electrons during internal thermalization of the electronic subsystem. This leads to non-equilibrium (non-thermal) spin excitations in the electronic structure. Small band shifts of about 40\,meV are attributed to the breakdown of spin-orbit coupling between hybridized bands upon ultrafast spin mixing. Since the bands show strong spin-orbit coupling, angular momentum can be dissipated transferring spin to orbital angular momentum. Remarkably the  $\Delta_1^\downarrow$\,-like band shows a three times larger decay-amplitude compared to the $\Delta_4^\downarrow$\,-like band. The non-equilibrium (non-thermal) dynamics within the spin system is attributed to the different phase space available for inelastic electron-magnon scattering.  This leads to significantly different spin mixing of both bands within the first picosecond after optical excitation, until thermal equilibrium between electron, phonon and spin systems is reached. In line with a recent theoretical study, we conclude that ultrafast demagnetization proceeds through fs generation of non-thermal magnons leading to band-specific spin mixing in the valence electronic structure \cite{Weissenhofer2025}.  
Our results are compatible with calculations and measurements of hot electron lifetimes in iron, where magnon contributions are particularly strong among the elemental $3d$ ferromagnets \cite{Zhukov2004,Schmidt2010,Weinelt2025}.
As already mentioned in the introduction, there are two time-resolved photoemission studies investigating the transient spin polarization of cobalt and iron in normal emission for $h\nu = 21 - 22$\,eV \cite{Eich2017,Gort2018}.  For Co/Cu(001), Eich and coworkers analyzed spin-resolved data comparing longitudinal and transvers spin excitations denoted as Stoner and band mirroring behavior, respectively. 
Assuming the same response of all bands with binding energy $E_{\rm{B}} > 0.5$\,eV they found best agreement for band mirroring and a rigid band structure. Furthermore, the cobalt spin-polarization decays with a time constant of 42\,fs at $E_{\rm{B}} = 2.3$\,eV \cite{Eich2017}. For Fe/W(110), Gort and coworkers found different response times of the spin polarization of 60\,fs close to $E_{\rm{F}}$ and of 450\,fs at $E_{\rm{B}} = 2.0$\,eV but comparable depolarization of $\sim 20$\,\%. Here the spin polarization started to recover around 1\,ps pump-probe delay \cite{Gort2018}.  Both results confirm the dominance of transverse spin excitations. The results on iron suggest non-equilibrium dynamics in the spin system, but by comparing optically pumped and non-pumped regions of the band structure. 
We note that with 10\,kHz repetition rate of the HHG source these are extremely demanding tr-ARPES experiments. To probe larger parts of the band structure will require higher repetition rates in combination with spin-resolved momentum microscopy \cite{Tusche2013}. 
In comparison, MLD in photoemission is a powerful, more simple but yet rarely employed tool to investigate spin dynamics. By studying exchange splitting and magnetic linear dichroism, we are able to distinguish between longitudinal and transverse spin excitations and show that ultrafast non-equilibrium magnon emission is crucial for Fe, confirming recent theoretical predictions \cite{Weissenhofer2025}. 
Generally, we expect that ultrafast magnon generation has significant impact on a number of phenomena in ultrafast spintronics such as the generation of THz radiation in Fe$|$heavy metal spin emitters \cite{Rouzegar2022}, the element-specific response times in alloys containing $3d$ elements \cite{KorffSchmising2024}, and  the angular momentum transfer in ultrafast magnetization switching of FeCoGd compounds \cite{Beens2019,Steinbach2024}.  

\begin{acknowledgments}
We acknowledge funding by the Deutsche Forschungsgemeinschaft (DFG, German Research Foundation) - Project-ID 328545488 CRC/TRR 227 \textit{Ultrafast Spin Dynamics}, project A01.
\end{acknowledgments}


\begin{thebibliography}{30}%
	\makeatletter
	\providecommand \@ifxundefined [1]{%
		\@ifx{#1\undefined}
	}%
	\providecommand \@ifnum [1]{%
		\ifnum #1\expandafter \@firstoftwo
		\else \expandafter \@secondoftwo
		\fi
	}%
	\providecommand \@ifx [1]{%
		\ifx #1\expandafter \@firstoftwo
		\else \expandafter \@secondoftwo
		\fi
	}%
	\providecommand \natexlab [1]{#1}%
	\providecommand \enquote  [1]{``#1''}%
	\providecommand \bibnamefont  [1]{#1}%
	\providecommand \bibfnamefont [1]{#1}%
	\providecommand \citenamefont [1]{#1}%
	\providecommand \href@noop [0]{\@secondoftwo}%
	\providecommand \href [0]{\begingroup \@sanitize@url \@href}%
	\providecommand \@href[1]{\@@startlink{#1}\@@href}%
	\providecommand \@@href[1]{\endgroup#1\@@endlink}%
	\providecommand \@sanitize@url [0]{\catcode `\\12\catcode `\$12\catcode
		`\&12\catcode `\#12\catcode `\^12\catcode `\_12\catcode `\%12\relax}%
	\providecommand \@@startlink[1]{}%
	\providecommand \@@endlink[0]{}%
	\providecommand \url  [0]{\begingroup\@sanitize@url \@url }%
	\providecommand \@url [1]{\endgroup\@href {#1}{\urlprefix }}%
	\providecommand \urlprefix  [0]{URL }%
	\providecommand \Eprint [0]{\href }%
	\providecommand \doibase [0]{https://doi.org/}%
	\providecommand \selectlanguage [0]{\@gobble}%
	\providecommand \bibinfo  [0]{\@secondoftwo}%
	\providecommand \bibfield  [0]{\@secondoftwo}%
	\providecommand \translation [1]{[#1]}%
	\providecommand \BibitemOpen [0]{}%
	\providecommand \bibitemStop [0]{}%
	\providecommand \bibitemNoStop [0]{.\EOS\space}%
	\providecommand \EOS [0]{\spacefactor3000\relax}%
	\providecommand \BibitemShut  [1]{\csname bibitem#1\endcsname}%
	\let\auto@bib@innerbib\@empty
	\bibitem [{\citenamefont {Wei{\ss}enhofer}\ and\ \citenamefont
		{Oppeneer}(2025)}]{Weissenhofer2025}%
	\BibitemOpen
	\bibfield  {author} {\bibinfo {author} {\bibfnamefont {M.}~\bibnamefont
			{Wei{\ss}enhofer}}\ and\ \bibinfo {author} {\bibfnamefont {P.~M.}\
			\bibnamefont {Oppeneer}},\ }\bibfield  {title} {\bibinfo {title} {Ultrafast
			demagnetization through femtosecond generation of non-thermal magnons},\
	}\href {https://doi.org/https://doi.org/10.1002/apxr.202300103} {\bibfield
		{journal} {\bibinfo  {journal} {Advanced Physics Research}\ }\textbf
		{\bibinfo {volume} {2300103}} (\bibinfo {year} {2025})}\BibitemShut {NoStop}%
	\bibitem [{\citenamefont {Rhie}\ \emph {et~al.}(2003)\citenamefont {Rhie},
		\citenamefont {D\"urr},\ and\ \citenamefont {Eberhardt}}]{Rhie2003}%
	\BibitemOpen
	\bibfield  {author} {\bibinfo {author} {\bibfnamefont {H.-S.}\ \bibnamefont
			{Rhie}}, \bibinfo {author} {\bibfnamefont {H.~A.}\ \bibnamefont {D\"urr}},\
		and\ \bibinfo {author} {\bibfnamefont {W.}~\bibnamefont {Eberhardt}},\
	}\bibfield  {title} {\bibinfo {title} {Femtosecond electron and spin dynamics
			in $\mathrm{N}\mathrm{i}/\mathrm{W}(110)$ films},\ }\href
	{https://doi.org/10.1103/PhysRevLett.90.247201} {\bibfield  {journal}
		{\bibinfo  {journal} {Phys. Rev. Lett.}\ }\textbf {\bibinfo {volume} {90}},\
		\bibinfo {pages} {247201} (\bibinfo {year} {2003})}\BibitemShut {NoStop}%
	\bibitem [{\citenamefont {Koopmans}\ \emph {et~al.}(2010)\citenamefont
		{Koopmans}, \citenamefont {Malinowski}, \citenamefont {Dalla~Longa},
		\citenamefont {Steiauf}, \citenamefont {F{\"a}hnle}, \citenamefont {Roth},
		\citenamefont {Cinchetti},\ and\ \citenamefont {Aeschlimann}}]{Koopmans2010}%
	\BibitemOpen
	\bibfield  {author} {\bibinfo {author} {\bibfnamefont {B.}~\bibnamefont
			{Koopmans}}, \bibinfo {author} {\bibfnamefont {G.}~\bibnamefont
			{Malinowski}}, \bibinfo {author} {\bibfnamefont {F.}~\bibnamefont
			{Dalla~Longa}}, \bibinfo {author} {\bibfnamefont {D.}~\bibnamefont
			{Steiauf}}, \bibinfo {author} {\bibfnamefont {M.}~\bibnamefont {F{\"a}hnle}},
		\bibinfo {author} {\bibfnamefont {T.}~\bibnamefont {Roth}}, \bibinfo {author}
		{\bibfnamefont {C.}~\bibnamefont {Cinchetti}},\ and\ \bibinfo {author}
		{\bibfnamefont {M.}~\bibnamefont {Aeschlimann}},\ }\bibfield  {title}
	{\bibinfo {title} {Explaining the paradoxical diversity of ultrafast
			laser-induced demagnetization},\ }\href
	{https://doi.org/doi:10.1038/nmat2593} {\bibfield  {journal} {\bibinfo
			{journal} {Nature Mater.}\ }\textbf {\bibinfo {volume} {9}},\ \bibinfo
		{pages} {259} (\bibinfo {year} {2010})}\BibitemShut {NoStop}%
	\bibitem [{\citenamefont {Mueller}\ \emph {et~al.}(2013)\citenamefont
		{Mueller}, \citenamefont {Baral}, \citenamefont {Vollmar}, \citenamefont
		{Cinchetti}, \citenamefont {Aeschlimann}, \citenamefont {Schneider},\ and\
		\citenamefont {Rethfeld}}]{Mueller2013}%
	\BibitemOpen
	\bibfield  {author} {\bibinfo {author} {\bibfnamefont {B.~Y.}\ \bibnamefont
			{Mueller}}, \bibinfo {author} {\bibfnamefont {A.}~\bibnamefont {Baral}},
		\bibinfo {author} {\bibfnamefont {S.}~\bibnamefont {Vollmar}}, \bibinfo
		{author} {\bibfnamefont {M.}~\bibnamefont {Cinchetti}}, \bibinfo {author}
		{\bibfnamefont {M.}~\bibnamefont {Aeschlimann}}, \bibinfo {author}
		{\bibfnamefont {H.~C.}\ \bibnamefont {Schneider}},\ and\ \bibinfo {author}
		{\bibfnamefont {B.}~\bibnamefont {Rethfeld}},\ }\bibfield  {title} {\bibinfo
		{title} {Feedback effect during ultrafast demagnetization dynamics in
			ferromagnets},\ }\href {https://doi.org/10.1103/PhysRevLett.111.167204}
	{\bibfield  {journal} {\bibinfo  {journal} {Phys. Rev. Lett.}\ }\textbf
		{\bibinfo {volume} {111}},\ \bibinfo {pages} {167204} (\bibinfo {year}
		{2013})}\BibitemShut {NoStop}%
	\bibitem [{\citenamefont {Griepe}\ and\ \citenamefont
		{Atxitia}(2023)}]{Griepe2023}%
	\BibitemOpen
	\bibfield  {author} {\bibinfo {author} {\bibfnamefont {T.}~\bibnamefont
			{Griepe}}\ and\ \bibinfo {author} {\bibfnamefont {U.}~\bibnamefont
			{Atxitia}},\ }\bibfield  {title} {\bibinfo {title} {Evidence of
			electron-phonon mediated spin flip as driving mechanism for ultrafast
			magnetization dynamics in $3d$ ferromagnets},\ }\href
	{https://doi.org/10.1103/PhysRevB.107.L100407} {\bibfield  {journal}
		{\bibinfo  {journal} {Phys. Rev. B}\ }\textbf {\bibinfo {volume} {107}},\
		\bibinfo {pages} {L100407} (\bibinfo {year} {2023})}\BibitemShut {NoStop}%
	\bibitem [{\citenamefont {Carpene}\ \emph {et~al.}(2008)\citenamefont
		{Carpene}, \citenamefont {Mancini}, \citenamefont {Dallera}, \citenamefont
		{Brenna}, \citenamefont {Puppin},\ and\ \citenamefont
		{De~Silvestri}}]{Carpene2008}%
	\BibitemOpen
	\bibfield  {author} {\bibinfo {author} {\bibfnamefont {E.}~\bibnamefont
			{Carpene}}, \bibinfo {author} {\bibfnamefont {E.}~\bibnamefont {Mancini}},
		\bibinfo {author} {\bibfnamefont {C.}~\bibnamefont {Dallera}}, \bibinfo
		{author} {\bibfnamefont {M.}~\bibnamefont {Brenna}}, \bibinfo {author}
		{\bibfnamefont {E.}~\bibnamefont {Puppin}},\ and\ \bibinfo {author}
		{\bibfnamefont {S.}~\bibnamefont {De~Silvestri}},\ }\bibfield  {title}
	{\bibinfo {title} {Dynamics of electron-magnon interaction and ultrafast
			demagnetization in thin iron films},\ }\href
	{https://doi.org/10.1103/PhysRevB.78.174422} {\bibfield  {journal} {\bibinfo
			{journal} {Phys. Rev. B}\ }\textbf {\bibinfo {volume} {78}},\ \bibinfo
		{pages} {174422} (\bibinfo {year} {2008})}\BibitemShut {NoStop}%
	\bibitem [{\citenamefont {Schmidt}\ \emph {et~al.}(2010)\citenamefont
		{Schmidt}, \citenamefont {Pickel}, \citenamefont {Donath}, \citenamefont
		{Buczek}, \citenamefont {Ernst}, \citenamefont {Zhukov}, \citenamefont
		{Echenique}, \citenamefont {Sandratskii}, \citenamefont {Chulkov},\ and\
		\citenamefont {Weinelt}}]{Schmidt2010}%
	\BibitemOpen
	\bibfield  {author} {\bibinfo {author} {\bibfnamefont {A.~B.}\ \bibnamefont
			{Schmidt}}, \bibinfo {author} {\bibfnamefont {M.}~\bibnamefont {Pickel}},
		\bibinfo {author} {\bibfnamefont {M.}~\bibnamefont {Donath}}, \bibinfo
		{author} {\bibfnamefont {P.}~\bibnamefont {Buczek}}, \bibinfo {author}
		{\bibfnamefont {A.}~\bibnamefont {Ernst}}, \bibinfo {author} {\bibfnamefont
			{V.~P.}\ \bibnamefont {Zhukov}}, \bibinfo {author} {\bibfnamefont {P.~M.}\
			\bibnamefont {Echenique}}, \bibinfo {author} {\bibfnamefont {L.~M.}\
			\bibnamefont {Sandratskii}}, \bibinfo {author} {\bibfnamefont {E.~V.}\
			\bibnamefont {Chulkov}},\ and\ \bibinfo {author} {\bibfnamefont
			{M.}~\bibnamefont {Weinelt}},\ }\bibfield  {title} {\bibinfo {title}
		{Ultrafast magnon generation in an {Fe} film on {Cu(100)}},\ }\href
	{https://doi.org/10.1103/PhysRevLett.105.197401} {\bibfield  {journal}
		{\bibinfo  {journal} {Phys. Rev. Lett.}\ }\textbf {\bibinfo {volume} {105}},\
		\bibinfo {pages} {197401} (\bibinfo {year} {2010})}\BibitemShut {NoStop}%
	\bibitem [{\citenamefont {Carpene}\ \emph {et~al.}(2015)\citenamefont
		{Carpene}, \citenamefont {Hedayat}, \citenamefont {Boschini},\ and\
		\citenamefont {Dallera}}]{Carpene2015}%
	\BibitemOpen
	\bibfield  {author} {\bibinfo {author} {\bibfnamefont {E.}~\bibnamefont
			{Carpene}}, \bibinfo {author} {\bibfnamefont {H.}~\bibnamefont {Hedayat}},
		\bibinfo {author} {\bibfnamefont {F.}~\bibnamefont {Boschini}},\ and\
		\bibinfo {author} {\bibfnamefont {C.}~\bibnamefont {Dallera}},\ }\bibfield
	{title} {\bibinfo {title} {Ultrafast demagnetization of metals: Collapsed
			exchange versus collective excitations},\ }\href
	{https://doi.org/10.1103/PhysRevB.91.174414} {\bibfield  {journal} {\bibinfo
			{journal} {Phys. Rev. B}\ }\textbf {\bibinfo {volume} {91}},\ \bibinfo
		{pages} {174414} (\bibinfo {year} {2015})}\BibitemShut {NoStop}%
	\bibitem [{\citenamefont {Eich}\ \emph {et~al.}(2017)\citenamefont {Eich},
		\citenamefont {Plötzing}, \citenamefont {Rollinger}, \citenamefont
		{Emmerich}, \citenamefont {Adam}, \citenamefont {Chen}, \citenamefont
		{Kapteyn}, \citenamefont {Murnane}, \citenamefont {Plucinski}, \citenamefont
		{Steil}, \citenamefont {Stadtmüller}, \citenamefont {Cinchetti},
		\citenamefont {Aeschlimann}, \citenamefont {Schneider},\ and\ \citenamefont
		{Mathias}}]{Eich2017}%
	\BibitemOpen
	\bibfield  {author} {\bibinfo {author} {\bibfnamefont {S.}~\bibnamefont
			{Eich}}, \bibinfo {author} {\bibfnamefont {M.}~\bibnamefont {Plötzing}},
		\bibinfo {author} {\bibfnamefont {M.}~\bibnamefont {Rollinger}}, \bibinfo
		{author} {\bibfnamefont {S.}~\bibnamefont {Emmerich}}, \bibinfo {author}
		{\bibfnamefont {R.}~\bibnamefont {Adam}}, \bibinfo {author} {\bibfnamefont
			{C.}~\bibnamefont {Chen}}, \bibinfo {author} {\bibfnamefont {H.~C.}\
			\bibnamefont {Kapteyn}}, \bibinfo {author} {\bibfnamefont {M.~M.}\
			\bibnamefont {Murnane}}, \bibinfo {author} {\bibfnamefont {L.}~\bibnamefont
			{Plucinski}}, \bibinfo {author} {\bibfnamefont {D.}~\bibnamefont {Steil}},
		\bibinfo {author} {\bibfnamefont {B.}~\bibnamefont {Stadtmüller}}, \bibinfo
		{author} {\bibfnamefont {M.}~\bibnamefont {Cinchetti}}, \bibinfo {author}
		{\bibfnamefont {M.}~\bibnamefont {Aeschlimann}}, \bibinfo {author}
		{\bibfnamefont {C.~M.}\ \bibnamefont {Schneider}},\ and\ \bibinfo {author}
		{\bibfnamefont {S.}~\bibnamefont {Mathias}},\ }\bibfield  {title} {\bibinfo
		{title} {Band structure evolution during the ultrafast
			ferromagnetic-paramagnetic phase transition in cobalt},\ }\href
	{https://doi.org/10.1126/sciadv.1602094} {\bibfield  {journal} {\bibinfo
			{journal} {Sci. Adv.}\ }\textbf {\bibinfo {volume} {3}},\ \bibinfo {pages}
		{e1602094} (\bibinfo {year} {2017})}\BibitemShut {NoStop}%
	\bibitem [{\citenamefont {Gort}\ \emph {et~al.}(2018)\citenamefont {Gort},
		\citenamefont {Bühlmann}, \citenamefont {Däster}, \citenamefont
		{Salvatella}, \citenamefont {Hartmann}, \citenamefont {Zemp}, \citenamefont
		{Holenstein}, \citenamefont {Stieger}, \citenamefont {Fognini}, \citenamefont
		{Michlmayr}, \citenamefont {Bähler}, \citenamefont {Vaterlaus},\ and\
		\citenamefont {Acremann}}]{Gort2018}%
	\BibitemOpen
	\bibfield  {author} {\bibinfo {author} {\bibfnamefont {R.}~\bibnamefont
			{Gort}}, \bibinfo {author} {\bibfnamefont {K.}~\bibnamefont {Bühlmann}},
		\bibinfo {author} {\bibfnamefont {S.}~\bibnamefont {Däster}}, \bibinfo
		{author} {\bibfnamefont {G.}~\bibnamefont {Salvatella}}, \bibinfo {author}
		{\bibfnamefont {N.}~\bibnamefont {Hartmann}}, \bibinfo {author}
		{\bibfnamefont {Y.}~\bibnamefont {Zemp}}, \bibinfo {author} {\bibfnamefont
			{S.}~\bibnamefont {Holenstein}}, \bibinfo {author} {\bibfnamefont
			{C.}~\bibnamefont {Stieger}}, \bibinfo {author} {\bibfnamefont
			{A.}~\bibnamefont {Fognini}}, \bibinfo {author} {\bibfnamefont
			{T.}~\bibnamefont {Michlmayr}}, \bibinfo {author} {\bibfnamefont
			{T.}~\bibnamefont {Bähler}}, \bibinfo {author} {\bibfnamefont
			{A.}~\bibnamefont {Vaterlaus}},\ and\ \bibinfo {author} {\bibfnamefont
			{Y.}~\bibnamefont {Acremann}},\ }\bibfield  {title} {\bibinfo {title} {Early
			stages of ultrafast spin dynamics in a $3d$ ferromagnet},\ }\href
	{https://doi.org/10.1103/PhysRevLett.121.087206} {\bibfield  {journal}
		{\bibinfo  {journal} {Phys. Rev. Lett.}\ }\textbf {\bibinfo {volume} {121}},\
		\bibinfo {pages} {087206} (\bibinfo {year} {2018})}\BibitemShut {NoStop}%
	\bibitem [{\citenamefont {Tengdin}\ \emph {et~al.}(2018)\citenamefont
		{Tengdin}, \citenamefont {You}, \citenamefont {Chen}, \citenamefont {Shi},
		\citenamefont {Zusin}, \citenamefont {Zhang}, \citenamefont {Gentry},
		\citenamefont {Blonsky}, \citenamefont {Keller}, \citenamefont {Oppeneer},
		\citenamefont {Kapteyn}, \citenamefont {Tao},\ and\ \citenamefont
		{Murnane}}]{Tengdin2018}%
	\BibitemOpen
	\bibfield  {author} {\bibinfo {author} {\bibfnamefont {P.}~\bibnamefont
			{Tengdin}}, \bibinfo {author} {\bibfnamefont {W.}~\bibnamefont {You}},
		\bibinfo {author} {\bibfnamefont {C.}~\bibnamefont {Chen}}, \bibinfo {author}
		{\bibfnamefont {X.}~\bibnamefont {Shi}}, \bibinfo {author} {\bibfnamefont
			{D.}~\bibnamefont {Zusin}}, \bibinfo {author} {\bibfnamefont
			{Y.}~\bibnamefont {Zhang}}, \bibinfo {author} {\bibfnamefont
			{C.}~\bibnamefont {Gentry}}, \bibinfo {author} {\bibfnamefont
			{A.}~\bibnamefont {Blonsky}}, \bibinfo {author} {\bibfnamefont
			{M.}~\bibnamefont {Keller}}, \bibinfo {author} {\bibfnamefont {P.~M.}\
			\bibnamefont {Oppeneer}}, \bibinfo {author} {\bibfnamefont {H.~C.}\
			\bibnamefont {Kapteyn}}, \bibinfo {author} {\bibfnamefont {Z.}~\bibnamefont
			{Tao}},\ and\ \bibinfo {author} {\bibfnamefont {M.~M.}\ \bibnamefont
			{Murnane}},\ }\bibfield  {title} {\bibinfo {title} {Critical behavior within
			20 fs drives the out-of-equilibrium laser-induced magnetic phase transition
			in nickel},\ }\href {https://doi.org/10.1126/sciadv.aap9744} {\bibfield
		{journal} {\bibinfo  {journal} {Science Advances}\ }\textbf {\bibinfo
			{volume} {4}},\ \bibinfo {pages} {eaap9744} (\bibinfo {year}
		{2018})}\BibitemShut {NoStop}%
	\bibitem [{\citenamefont {S\'anchez-Barriga}\ \emph {et~al.}(2012)\citenamefont
		{S\'anchez-Barriga}, \citenamefont {Braun}, \citenamefont {Min\'ar},
		\citenamefont {Di~Marco}, \citenamefont {Varykhalov}, \citenamefont {Rader},
		\citenamefont {Boni}, \citenamefont {Bellini}, \citenamefont {Manghi},
		\citenamefont {Ebert}, \citenamefont {Katsnelson}, \citenamefont
		{Lichtenstein}, \citenamefont {Eriksson}, \citenamefont {Eberhardt},
		\citenamefont {D\"urr},\ and\ \citenamefont {Fink}}]{Sanchez-Barriga2012}%
	\BibitemOpen
	\bibfield  {author} {\bibinfo {author} {\bibfnamefont {J.}~\bibnamefont
			{S\'anchez-Barriga}}, \bibinfo {author} {\bibfnamefont {J.}~\bibnamefont
			{Braun}}, \bibinfo {author} {\bibfnamefont {J.}~\bibnamefont {Min\'ar}},
		\bibinfo {author} {\bibfnamefont {I.}~\bibnamefont {Di~Marco}}, \bibinfo
		{author} {\bibfnamefont {A.}~\bibnamefont {Varykhalov}}, \bibinfo {author}
		{\bibfnamefont {O.}~\bibnamefont {Rader}}, \bibinfo {author} {\bibfnamefont
			{V.}~\bibnamefont {Boni}}, \bibinfo {author} {\bibfnamefont {V.}~\bibnamefont
			{Bellini}}, \bibinfo {author} {\bibfnamefont {F.}~\bibnamefont {Manghi}},
		\bibinfo {author} {\bibfnamefont {H.}~\bibnamefont {Ebert}}, \bibinfo
		{author} {\bibfnamefont {M.~I.}\ \bibnamefont {Katsnelson}}, \bibinfo
		{author} {\bibfnamefont {A.~I.}\ \bibnamefont {Lichtenstein}}, \bibinfo
		{author} {\bibfnamefont {O.}~\bibnamefont {Eriksson}}, \bibinfo {author}
		{\bibfnamefont {W.}~\bibnamefont {Eberhardt}}, \bibinfo {author}
		{\bibfnamefont {H.~A.}\ \bibnamefont {D\"urr}},\ and\ \bibinfo {author}
		{\bibfnamefont {J.}~\bibnamefont {Fink}},\ }\bibfield  {title} {\bibinfo
		{title} {Effects of spin-dependent quasiparticle renormalization in {Fe},
			{Co}, and {Ni} photoemission spectra: {An} experimental and theoretical
			study},\ }\href {https://doi.org/10.1103/PhysRevB.85.205109} {\bibfield
		{journal} {\bibinfo  {journal} {Phys. Rev. B}\ }\textbf {\bibinfo {volume}
			{85}},\ \bibinfo {pages} {205109} (\bibinfo {year} {2012})}\BibitemShut
	{NoStop}%
	\bibitem [{\citenamefont {Tusche}\ \emph {et~al.}(2018)\citenamefont {Tusche},
		\citenamefont {Ellguth}, \citenamefont {Feyer}, \citenamefont {Krasyuk},
		\citenamefont {Wiemann}, \citenamefont {Henk}, \citenamefont {Schneider},\
		and\ \citenamefont {Kirschner}}]{Tusche2018}%
	\BibitemOpen
	\bibfield  {author} {\bibinfo {author} {\bibfnamefont {C.}~\bibnamefont
			{Tusche}}, \bibinfo {author} {\bibfnamefont {M.}~\bibnamefont {Ellguth}},
		\bibinfo {author} {\bibfnamefont {V.}~\bibnamefont {Feyer}}, \bibinfo
		{author} {\bibfnamefont {A.}~\bibnamefont {Krasyuk}}, \bibinfo {author}
		{\bibfnamefont {C.}~\bibnamefont {Wiemann}}, \bibinfo {author} {\bibfnamefont
			{J.}~\bibnamefont {Henk}}, \bibinfo {author} {\bibfnamefont {C.~M.}\
			\bibnamefont {Schneider}},\ and\ \bibinfo {author} {\bibfnamefont
			{J.}~\bibnamefont {Kirschner}},\ }\bibfield  {title} {\bibinfo {title}
		{Nonlocal electron correlations in an itinerant ferromagnet},\ }\href
	{https://doi.org/10.1038/s41467-018-05960-5} {\bibfield  {journal} {\bibinfo
			{journal} {Nat. Commun.}\ }\textbf {\bibinfo {volume} {9}},\ \bibinfo {pages}
		{3727} (\bibinfo {year} {2018})}\BibitemShut {NoStop}%
	\bibitem [{\citenamefont {Greber}\ \emph {et~al.}(1997)\citenamefont {Greber},
		\citenamefont {Kreutz},\ and\ \citenamefont {Osterwalder}}]{Greber1997}%
	\BibitemOpen
	\bibfield  {author} {\bibinfo {author} {\bibfnamefont {T.}~\bibnamefont
			{Greber}}, \bibinfo {author} {\bibfnamefont {T.~J.}\ \bibnamefont {Kreutz}},\
		and\ \bibinfo {author} {\bibfnamefont {J.}~\bibnamefont {Osterwalder}},\
	}\bibfield  {title} {\bibinfo {title} {Photoemission above the fermi level:
			The top of the minority $\mathit{d}$ band in nickel},\ }\href
	{https://doi.org/10.1103/PhysRevLett.79.4465} {\bibfield  {journal} {\bibinfo
			{journal} {Phys. Rev. Lett.}\ }\textbf {\bibinfo {volume} {79}},\ \bibinfo
		{pages} {4465} (\bibinfo {year} {1997})}\BibitemShut {NoStop}%
	\bibitem [{\citenamefont {Bovensiepen}(2007)}]{Bovensiepen2007}%
	\BibitemOpen
	\bibfield  {author} {\bibinfo {author} {\bibfnamefont {U.}~\bibnamefont
			{Bovensiepen}},\ }\bibfield  {title} {\bibinfo {title} {Coherent and
			incoherent excitations of the {Gd(0001)} surface on ultrafast timescales},\
	}\href {https://doi.org/10.1088/0953-8984/19/8/083201} {\bibfield  {journal}
		{\bibinfo  {journal} {Journal of Physics: Condensed Matter}\ }\textbf
		{\bibinfo {volume} {19}},\ \bibinfo {pages} {083201} (\bibinfo {year}
		{2007})}\BibitemShut {NoStop}%
	\bibitem [{\citenamefont {Aeschlimann}\ \emph {et~al.}(2025)\citenamefont
		{Aeschlimann}, \citenamefont {Bange}, \citenamefont {Bauer}, \citenamefont
		{Bovensiepen}, \citenamefont {Elmers}, \citenamefont {Fauster}, \citenamefont
		{Gierster}, \citenamefont {Höfer}, \citenamefont {Huber}, \citenamefont
		{Li}, \citenamefont {Li}, \citenamefont {Mathias}, \citenamefont
		{Morgenstern}, \citenamefont {Petek}, \citenamefont {Reutzel}, \citenamefont
		{Rossnagel}, \citenamefont {Schönhense}, \citenamefont {Scholz},
		\citenamefont {Stadtmüller}, \citenamefont {Stähler}, \citenamefont {Tan},
		\citenamefont {Wang}, \citenamefont {Wang},\ and\ \citenamefont
		{Weinelt}}]{Weinelt2025}%
	\BibitemOpen
	\bibfield  {author} {\bibinfo {author} {\bibfnamefont {M.}~\bibnamefont
			{Aeschlimann}}, \bibinfo {author} {\bibfnamefont {J.~P.}\ \bibnamefont
			{Bange}}, \bibinfo {author} {\bibfnamefont {M.}~\bibnamefont {Bauer}},
		\bibinfo {author} {\bibfnamefont {U.}~\bibnamefont {Bovensiepen}}, \bibinfo
		{author} {\bibfnamefont {H.-J.}\ \bibnamefont {Elmers}}, \bibinfo {author}
		{\bibfnamefont {T.}~\bibnamefont {Fauster}}, \bibinfo {author} {\bibfnamefont
			{L.}~\bibnamefont {Gierster}}, \bibinfo {author} {\bibfnamefont
			{U.}~\bibnamefont {Höfer}}, \bibinfo {author} {\bibfnamefont
			{R.}~\bibnamefont {Huber}}, \bibinfo {author} {\bibfnamefont
			{A.}~\bibnamefont {Li}}, \bibinfo {author} {\bibfnamefont {X.}~\bibnamefont
			{Li}}, \bibinfo {author} {\bibfnamefont {S.}~\bibnamefont {Mathias}},
		\bibinfo {author} {\bibfnamefont {K.}~\bibnamefont {Morgenstern}}, \bibinfo
		{author} {\bibfnamefont {H.}~\bibnamefont {Petek}}, \bibinfo {author}
		{\bibfnamefont {M.}~\bibnamefont {Reutzel}}, \bibinfo {author} {\bibfnamefont
			{K.}~\bibnamefont {Rossnagel}}, \bibinfo {author} {\bibfnamefont
			{G.}~\bibnamefont {Schönhense}}, \bibinfo {author} {\bibfnamefont
			{M.}~\bibnamefont {Scholz}}, \bibinfo {author} {\bibfnamefont
			{B.}~\bibnamefont {Stadtmüller}}, \bibinfo {author} {\bibfnamefont
			{J.}~\bibnamefont {Stähler}}, \bibinfo {author} {\bibfnamefont
			{S.}~\bibnamefont {Tan}}, \bibinfo {author} {\bibfnamefont {B.}~\bibnamefont
			{Wang}}, \bibinfo {author} {\bibfnamefont {Z.}~\bibnamefont {Wang}},\ and\
		\bibinfo {author} {\bibfnamefont {M.}~\bibnamefont {Weinelt}},\ }\bibfield
	{title} {\bibinfo {title} {Time-resolved photoelectron spectroscopy at
			surfaces},\ }\href
	{https://doi.org/https://doi.org/10.1016/j.susc.2024.122631} {\bibfield
		{journal} {\bibinfo  {journal} {Surf. Sci.}\ }\textbf {\bibinfo {volume}
			{753}},\ \bibinfo {pages} {122631} (\bibinfo {year} {2025})},\ \bibinfo
	{note} {{Chapter 8}, Ultrafast Spin Dynamics probed by tr-ARPES}\BibitemShut
	{NoStop}%
	\bibitem [{\citenamefont {Zhukov}\ \emph {et~al.}(2004)\citenamefont {Zhukov},
		\citenamefont {Chulkov},\ and\ \citenamefont {Echenique}}]{Zhukov2004}%
	\BibitemOpen
	\bibfield  {author} {\bibinfo {author} {\bibfnamefont {V.~P.}\ \bibnamefont
			{Zhukov}}, \bibinfo {author} {\bibfnamefont {E.~V.}\ \bibnamefont
			{Chulkov}},\ and\ \bibinfo {author} {\bibfnamefont {P.~M.}\ \bibnamefont
			{Echenique}},\ }\bibfield  {title} {\bibinfo {title} {Lifetimes of excited
			electrons in fe and ni: First-principles gw and the $t$-matrix theory},\
	}\href {https://doi.org/10.1103/PhysRevLett.93.096401} {\bibfield  {journal}
		{\bibinfo  {journal} {Phys. Rev. Lett.}\ }\textbf {\bibinfo {volume} {93}},\
		\bibinfo {pages} {096401} (\bibinfo {year} {2004})}\BibitemShut {NoStop}%
	\bibitem [{\citenamefont {Frietsch}\ \emph {et~al.}(2013)\citenamefont
		{Frietsch}, \citenamefont {Carley}, \citenamefont {Döbrich}, \citenamefont
		{Gahl}, \citenamefont {Teichmann}, \citenamefont {Schwarzkopf}, \citenamefont
		{Wernet},\ and\ \citenamefont {Weinelt}}]{Frietsch2013}%
	\BibitemOpen
	\bibfield  {author} {\bibinfo {author} {\bibfnamefont {B.}~\bibnamefont
			{Frietsch}}, \bibinfo {author} {\bibfnamefont {R.}~\bibnamefont {Carley}},
		\bibinfo {author} {\bibfnamefont {K.}~\bibnamefont {Döbrich}}, \bibinfo
		{author} {\bibfnamefont {C.}~\bibnamefont {Gahl}}, \bibinfo {author}
		{\bibfnamefont {M.}~\bibnamefont {Teichmann}}, \bibinfo {author}
		{\bibfnamefont {O.}~\bibnamefont {Schwarzkopf}}, \bibinfo {author}
		{\bibfnamefont {P.}~\bibnamefont {Wernet}},\ and\ \bibinfo {author}
		{\bibfnamefont {M.}~\bibnamefont {Weinelt}},\ }\bibfield  {title} {\bibinfo
		{title} {A high-order harmonic generation apparatus for time- and
			angle-resolved photoelectron spectroscopy},\ }\href
	{https://doi.org/10.1063/1.4812992} {\bibfield  {journal} {\bibinfo
			{journal} {Rev. of Sci. Instrum.}\ }\textbf {\bibinfo {volume} {84}},\
		\bibinfo {pages} {075106} (\bibinfo {year} {2013})}\BibitemShut {NoStop}%
	\bibitem [{\citenamefont {Gradmann}\ and\ \citenamefont
		{Waller}(1982)}]{Gradmann1982}%
	\BibitemOpen
	\bibfield  {author} {\bibinfo {author} {\bibfnamefont {U.}~\bibnamefont
			{Gradmann}}\ and\ \bibinfo {author} {\bibfnamefont {G.}~\bibnamefont
			{Waller}},\ }\bibfield  {title} {\bibinfo {title} {Periodic lattice
			distortions in epitaxial films of {Fe(110)} on {W(110)}},\ }\href
	{https://doi.org/10.1016/0039-6028(82)90363-6} {\bibfield  {journal}
		{\bibinfo  {journal} {Surf. Sci.}\ }\textbf {\bibinfo {volume} {116}},\
		\bibinfo {pages} {539} (\bibinfo {year} {1982})}\BibitemShut {NoStop}%
	\bibitem [{\citenamefont {Andres}\ \emph {et~al.}(2015)\citenamefont {Andres},
		\citenamefont {Christ}, \citenamefont {Gahl}, \citenamefont {Wietstruk},
		\citenamefont {Weinelt},\ and\ \citenamefont {Kirschner}}]{Andres2015}%
	\BibitemOpen
	\bibfield  {author} {\bibinfo {author} {\bibfnamefont {B.}~\bibnamefont
			{Andres}}, \bibinfo {author} {\bibfnamefont {M.}~\bibnamefont {Christ}},
		\bibinfo {author} {\bibfnamefont {C.}~\bibnamefont {Gahl}}, \bibinfo {author}
		{\bibfnamefont {M.}~\bibnamefont {Wietstruk}}, \bibinfo {author}
		{\bibfnamefont {M.}~\bibnamefont {Weinelt}},\ and\ \bibinfo {author}
		{\bibfnamefont {J.}~\bibnamefont {Kirschner}},\ }\bibfield  {title} {\bibinfo
		{title} {Separating {Exchange} {Splitting} from {Spin} {Mixing} in
			{Gadolinium} by {Femtosecond} {Laser} {Excitation}},\ }\href
	{https://doi.org/10.1103/PhysRevLett.115.207404} {\bibfield  {journal}
		{\bibinfo  {journal} {Phys. Rev. Lett.}\ }\textbf {\bibinfo {volume} {115}},\
		\bibinfo {pages} {207404} (\bibinfo {year} {2015})}\BibitemShut {NoStop}%
	\bibitem [{\citenamefont {Schäfer}\ \emph {et~al.}(2005)\citenamefont
		{Schäfer}, \citenamefont {Hoinkis}, \citenamefont {Rotenberg}, \citenamefont
		{Blaha},\ and\ \citenamefont {Claessen}}]{Schaefer2005}%
	\BibitemOpen
	\bibfield  {author} {\bibinfo {author} {\bibfnamefont {J.}~\bibnamefont
			{Schäfer}}, \bibinfo {author} {\bibfnamefont {M.}~\bibnamefont {Hoinkis}},
		\bibinfo {author} {\bibfnamefont {E.}~\bibnamefont {Rotenberg}}, \bibinfo
		{author} {\bibfnamefont {P.}~\bibnamefont {Blaha}},\ and\ \bibinfo {author}
		{\bibfnamefont {R.}~\bibnamefont {Claessen}},\ }\bibfield  {title} {\bibinfo
		{title} {Fermi surface and electron correlation effects of ferromagnetic
			iron},\ }\href {https://doi.org/10.1103/PhysRevB.72.155115} {\bibfield
		{journal} {\bibinfo  {journal} {Phys. Rev. B}\ }\textbf {\bibinfo {volume}
			{72}},\ \bibinfo {pages} {155115} (\bibinfo {year} {2005})}\BibitemShut
	{NoStop}%
	\bibitem [{\citenamefont {Andres}\ \emph {et~al.}(2022)\citenamefont {Andres},
		\citenamefont {Weinelt}, \citenamefont {Ebert}, \citenamefont {Braun},
		\citenamefont {Aperis},\ and\ \citenamefont {Oppeneer}}]{Andres2022}%
	\BibitemOpen
	\bibfield  {author} {\bibinfo {author} {\bibfnamefont {B.}~\bibnamefont
			{Andres}}, \bibinfo {author} {\bibfnamefont {M.}~\bibnamefont {Weinelt}},
		\bibinfo {author} {\bibfnamefont {H.}~\bibnamefont {Ebert}}, \bibinfo
		{author} {\bibfnamefont {J.}~\bibnamefont {Braun}}, \bibinfo {author}
		{\bibfnamefont {A.}~\bibnamefont {Aperis}},\ and\ \bibinfo {author}
		{\bibfnamefont {P.~M.}\ \bibnamefont {Oppeneer}},\ }\bibfield  {title}
	{\bibinfo {title} {Strong momentum-dependent electron–magnon
			renormalization of a surface resonance on iron},\ }\href
	{https://doi.org/10.1063/5.0089688} {\bibfield  {journal} {\bibinfo
			{journal} {Applied Physics Letters}\ }\textbf {\bibinfo {volume} {120}},\
		\bibinfo {pages} {202404} (\bibinfo {year} {2022})}\BibitemShut {NoStop}%
	\bibitem [{\citenamefont {Rampe}\ \emph {et~al.}(1998)\citenamefont {Rampe},
		\citenamefont {G\"untherodt}, \citenamefont {Hartmann}, \citenamefont {Henk},
		\citenamefont {Scheunemann},\ and\ \citenamefont {Feder}}]{Rampe1998}%
	\BibitemOpen
	\bibfield  {author} {\bibinfo {author} {\bibfnamefont {A.}~\bibnamefont
			{Rampe}}, \bibinfo {author} {\bibfnamefont {G.}~\bibnamefont {G\"untherodt}},
		\bibinfo {author} {\bibfnamefont {D.}~\bibnamefont {Hartmann}}, \bibinfo
		{author} {\bibfnamefont {J.}~\bibnamefont {Henk}}, \bibinfo {author}
		{\bibfnamefont {T.}~\bibnamefont {Scheunemann}},\ and\ \bibinfo {author}
		{\bibfnamefont {R.}~\bibnamefont {Feder}},\ }\bibfield  {title} {\bibinfo
		{title} {Magnetic linear dichroism in valence-band photoemission:
			Experimental and theoretical study of {Fe(110)}},\ }\href
	{https://doi.org/10.1103/PhysRevB.57.14370} {\bibfield  {journal} {\bibinfo
			{journal} {Phys. Rev. B}\ }\textbf {\bibinfo {volume} {57}},\ \bibinfo
		{pages} {14370} (\bibinfo {year} {1998})}\BibitemShut {NoStop}%
	\bibitem [{Not()}]{Note1}%
	\BibitemOpen
	\href@noop {} {}\bibinfo {note} {We use the notation $\Delta$-like to
		indicate that we measure $15^{\circ}$ of $\Gamma$-H.}\BibitemShut {Stop}%
	\bibitem [{\citenamefont {Kuch}\ and\ \citenamefont
		{Schneider}(2001)}]{Kuch2001}%
	\BibitemOpen
	\bibfield  {author} {\bibinfo {author} {\bibfnamefont {W.}~\bibnamefont
			{Kuch}}\ and\ \bibinfo {author} {\bibfnamefont {C.~M.}\ \bibnamefont
			{Schneider}},\ }\bibfield  {title} {\bibinfo {title} {Magnetic dichroism in
			valence band photoemission},\ }\href
	{https://doi.org/10.1088/0034-4885/64/2/201} {\bibfield  {journal} {\bibinfo
			{journal} {Reports on Progress in Physics}\ }\textbf {\bibinfo {volume}
			{64}},\ \bibinfo {pages} {147} (\bibinfo {year} {2001})}\BibitemShut
	{NoStop}%
	\bibitem [{\citenamefont {Tusche}\ \emph {et~al.}(2013)\citenamefont {Tusche},
		\citenamefont {Ellguth}, \citenamefont {Krasyuk}, \citenamefont {Winkelmann},
		\citenamefont {Kutnyakhov}, \citenamefont {Lushchyk}, \citenamefont
		{Medjanik}, \citenamefont {Schönhense},\ and\ \citenamefont
		{Kirschner}}]{Tusche2013}%
	\BibitemOpen
	\bibfield  {author} {\bibinfo {author} {\bibfnamefont {C.}~\bibnamefont
			{Tusche}}, \bibinfo {author} {\bibfnamefont {M.}~\bibnamefont {Ellguth}},
		\bibinfo {author} {\bibfnamefont {A.}~\bibnamefont {Krasyuk}}, \bibinfo
		{author} {\bibfnamefont {A.}~\bibnamefont {Winkelmann}}, \bibinfo {author}
		{\bibfnamefont {D.}~\bibnamefont {Kutnyakhov}}, \bibinfo {author}
		{\bibfnamefont {P.}~\bibnamefont {Lushchyk}}, \bibinfo {author}
		{\bibfnamefont {K.}~\bibnamefont {Medjanik}}, \bibinfo {author}
		{\bibfnamefont {G.}~\bibnamefont {Schönhense}},\ and\ \bibinfo {author}
		{\bibfnamefont {J.}~\bibnamefont {Kirschner}},\ }\bibfield  {title} {\bibinfo
		{title} {Quantitative spin polarization analysis in photoelectron emission
			microscopy with an imaging spin filter},\ }\href
	{https://doi.org/10.1016/j.ultramic.2013.02.022} {\bibfield  {journal}
		{\bibinfo  {journal} {Ultramicroscopy}\ }\textbf {\bibinfo {volume} {130}},\
		\bibinfo {pages} {70} (\bibinfo {year} {2013})},\ \bibinfo {note} {eighth
		International Workshop on LEEM/PEEM}\BibitemShut {NoStop}%
	\bibitem [{\citenamefont {Rouzegar}\ \emph {et~al.}(2022)\citenamefont
		{Rouzegar}, \citenamefont {Brandt}, \citenamefont {N\'advorn\'{\i}k},
		\citenamefont {Reiss}, \citenamefont {Chekhov}, \citenamefont {Gueckstock},
		\citenamefont {In}, \citenamefont {Wolf}, \citenamefont {Seifert},
		\citenamefont {Brouwer}, \citenamefont {Woltersdorf},\ and\ \citenamefont
		{Kampfrath}}]{Rouzegar2022}%
	\BibitemOpen
	\bibfield  {author} {\bibinfo {author} {\bibfnamefont {R.}~\bibnamefont
			{Rouzegar}}, \bibinfo {author} {\bibfnamefont {L.}~\bibnamefont {Brandt}},
		\bibinfo {author} {\bibfnamefont {L.~c.~v.}\ \bibnamefont
			{N\'advorn\'{\i}k}}, \bibinfo {author} {\bibfnamefont {D.~A.}\ \bibnamefont
			{Reiss}}, \bibinfo {author} {\bibfnamefont {A.~L.}\ \bibnamefont {Chekhov}},
		\bibinfo {author} {\bibfnamefont {O.}~\bibnamefont {Gueckstock}}, \bibinfo
		{author} {\bibfnamefont {C.}~\bibnamefont {In}}, \bibinfo {author}
		{\bibfnamefont {M.}~\bibnamefont {Wolf}}, \bibinfo {author} {\bibfnamefont
			{T.~S.}\ \bibnamefont {Seifert}}, \bibinfo {author} {\bibfnamefont {P.~W.}\
			\bibnamefont {Brouwer}}, \bibinfo {author} {\bibfnamefont {G.}~\bibnamefont
			{Woltersdorf}},\ and\ \bibinfo {author} {\bibfnamefont {T.}~\bibnamefont
			{Kampfrath}},\ }\bibfield  {title} {\bibinfo {title} {Laser-induced terahertz
			spin transport in magnetic nanostructures arises from the same force as
			ultrafast demagnetization},\ }\href
	{https://doi.org/10.1103/PhysRevB.106.144427} {\bibfield  {journal} {\bibinfo
			{journal} {Phys. Rev. B}\ }\textbf {\bibinfo {volume} {106}},\ \bibinfo
		{pages} {144427} (\bibinfo {year} {2022})}\BibitemShut {NoStop}%
	\bibitem [{\citenamefont {von Korff~Schmising}\ \emph
		{et~al.}(2024)\citenamefont {von Korff~Schmising}, \citenamefont {Jana},
		\citenamefont {Z\"ulich}, \citenamefont {Sommer},\ and\ \citenamefont
		{Eisebitt}}]{KorffSchmising2024}%
	\BibitemOpen
	\bibfield  {author} {\bibinfo {author} {\bibfnamefont {C.}~\bibnamefont {von
				Korff~Schmising}}, \bibinfo {author} {\bibfnamefont {S.}~\bibnamefont
			{Jana}}, \bibinfo {author} {\bibfnamefont {O.}~\bibnamefont {Z\"ulich}},
		\bibinfo {author} {\bibfnamefont {D.}~\bibnamefont {Sommer}},\ and\ \bibinfo
		{author} {\bibfnamefont {S.}~\bibnamefont {Eisebitt}},\ }\bibfield  {title}
	{\bibinfo {title} {Direct versus indirect excitation of ultrafast
			magnetization dynamics in {FeNi} alloys},\ }\href
	{https://doi.org/10.1103/PhysRevResearch.6.013270} {\bibfield  {journal}
		{\bibinfo  {journal} {Phys. Rev. Res.}\ }\textbf {\bibinfo {volume} {6}},\
		\bibinfo {pages} {013270} (\bibinfo {year} {2024})}\BibitemShut {NoStop}%
	\bibitem [{\citenamefont {Beens}\ \emph {et~al.}(2019)\citenamefont {Beens},
		\citenamefont {Lalieu}, \citenamefont {Deenen}, \citenamefont {Duine},\ and\
		\citenamefont {Koopmans}}]{Beens2019}%
	\BibitemOpen
	\bibfield  {author} {\bibinfo {author} {\bibfnamefont {M.}~\bibnamefont
			{Beens}}, \bibinfo {author} {\bibfnamefont {M.~L.~M.}\ \bibnamefont
			{Lalieu}}, \bibinfo {author} {\bibfnamefont {A.~J.~M.}\ \bibnamefont
			{Deenen}}, \bibinfo {author} {\bibfnamefont {R.~A.}\ \bibnamefont {Duine}},\
		and\ \bibinfo {author} {\bibfnamefont {B.}~\bibnamefont {Koopmans}},\
	}\bibfield  {title} {\bibinfo {title} {Comparing all-optical switching in
			synthetic-ferrimagnetic multilayers and alloys},\ }\href
	{https://doi.org/10.1103/PhysRevB.100.220409} {\bibfield  {journal} {\bibinfo
			{journal} {Phys. Rev. B}\ }\textbf {\bibinfo {volume} {100}},\ \bibinfo
		{pages} {220409(R)} (\bibinfo {year} {2019})}\BibitemShut {NoStop}%
	\bibitem [{\citenamefont {Steinbach}\ \emph {et~al.}(2024)\citenamefont
		{Steinbach}, \citenamefont {Atxitia}, \citenamefont {Yao}, \citenamefont
		{Borchert}, \citenamefont {Engel}, \citenamefont {Bencivenga}, \citenamefont
		{Foglia}, \citenamefont {Mincigrucci}, \citenamefont {Pedersoli},
		\citenamefont {De~Angelis}, \citenamefont {Pancaldi}, \citenamefont
		{Fainozzi}, \citenamefont {Pelli~Cresi}, \citenamefont {Paltanin},
		\citenamefont {Capotondi}, \citenamefont {Masciovecchio}, \citenamefont
		{Eisebitt},\ and\ \citenamefont {Schmising}}]{Steinbach2024}%
	\BibitemOpen
	\bibfield  {author} {\bibinfo {author} {\bibfnamefont {F.}~\bibnamefont
			{Steinbach}}, \bibinfo {author} {\bibfnamefont {U.}~\bibnamefont {Atxitia}},
		\bibinfo {author} {\bibfnamefont {K.}~\bibnamefont {Yao}}, \bibinfo {author}
		{\bibfnamefont {M.}~\bibnamefont {Borchert}}, \bibinfo {author}
		{\bibfnamefont {D.}~\bibnamefont {Engel}}, \bibinfo {author} {\bibfnamefont
			{F.}~\bibnamefont {Bencivenga}}, \bibinfo {author} {\bibfnamefont
			{L.}~\bibnamefont {Foglia}}, \bibinfo {author} {\bibfnamefont
			{R.}~\bibnamefont {Mincigrucci}}, \bibinfo {author} {\bibfnamefont
			{E.}~\bibnamefont {Pedersoli}}, \bibinfo {author} {\bibfnamefont
			{D.}~\bibnamefont {De~Angelis}}, \bibinfo {author} {\bibfnamefont
			{M.}~\bibnamefont {Pancaldi}}, \bibinfo {author} {\bibfnamefont
			{D.}~\bibnamefont {Fainozzi}}, \bibinfo {author} {\bibfnamefont {J.~S.}\
			\bibnamefont {Pelli~Cresi}}, \bibinfo {author} {\bibfnamefont
			{E.}~\bibnamefont {Paltanin}}, \bibinfo {author} {\bibfnamefont
			{F.}~\bibnamefont {Capotondi}}, \bibinfo {author} {\bibfnamefont
			{C.}~\bibnamefont {Masciovecchio}}, \bibinfo {author} {\bibfnamefont
			{S.}~\bibnamefont {Eisebitt}},\ and\ \bibinfo {author} {\bibfnamefont
			{C.}~\bibnamefont {Schmising}},\ }\bibfield  {title} {\bibinfo {title}
		{Exploring the fundamental spatial limits of magnetic all-optical
			switching},\ }\href {https://doi.org/10.1021/acs.nanolett.4c00129} {\bibfield
		{journal} {\bibinfo  {journal} {Nano letters}\ }\textbf {\bibinfo {volume}
			{24}} (\bibinfo {year} {2024})}\BibitemShut {NoStop}%
\end{thebibliography}
%

\end{document}